\documentclass[prl,twocolumn,amsmath]{revtex4}
\usepackage{graphicx}
\usepackage{dcolumn}
\usepackage{bm}

\begin{document}

\title{High Bandwidth Atomic Magnetometery with \\
 Continuous Quantum Non-demolition Measurements}
\author{V. Shah, G. Vasilakis and M. V. Romalis}
\affiliation{Department of Physics,  Princeton University,
Princeton, New Jersey 08544, USA}

\begin{abstract}
We describe an experimental study of spin-projection noise in a high
sensitivity alkali-metal magnetometer. We demonstrate a four-fold
improvement in the measurement bandwidth of the magnetometer using
continuous quantum non-demolition (QND) measurements. Operating in
the scalar mode with a measurement volume of 2 cm$^{3}$ we achieve
magnetic field sensitivity of 22 fT/Hz$^{1/2}$ and a bandwidth of
1.9 kHz with a spin polarization of only 1\%. Our experimental
arrangement is naturally back-action evading and can be used to
realize sub-fT sensitivity with a highly polarized spin-squeezed
atomic vapor.
\end{abstract}

\maketitle



\affiliation{Department of Physics, \newline
Princeton University, Princeton,\newline
New Jersey 08544, USA}



The limits imposed by quantum mechanics on precision measurements
have been the subject of long-standing interest. They are
particularly important in atomic systems that form the basis of
leading frequency standards, magnetometers and inertial sensors. The
Heisenberg uncertainty principle imposes a limit on measurement
sensitivity with uncorrelated atoms known as the standard quantum
limit (SQL). One can improve upon this limit using spin squeezing
techniques \cite{Wineland}. However, it has been shown theoretically
that in the presence of a constant rate of decoherence spin
squeezing does not lead to a significant improvement of the long
term measurement sensitivity, but can be used to reduce the
measurement time \cite{Huelga,Andre,Auzinsh}. Such an increase in
the measurement bandwidth without loss of sensitivity is
particularly important for systems with long spin coherence times,
which can achieve the highest measurement resolution but often
require impractically-long interrogation times.

Here we study the limits imposed by quantum spin fluctuations in a
dense hot alkali-metal vapor used in sensitive atomic magnetometers
\cite{MagNatRev}. While the sensitivity of most previous atomic
magnetometers has been limited by photon shot noise or technical
noise, we investigate the regime limited by spin projection noise.
We show that in this regime the measurement bandwidth can be
increased using quantum-non-demolition paramagnetic Faraday rotation
measurements \cite{Takahashi} without loss of sensitivity. We point
out that the bandwidth increase can be realized with only a weak
squeezing condition $(\Delta J_x)^2<J/2$ for a system with total
spin $J$, and demonstrate experimentally an increase in the
magnetometer bandwidth from 420 Hz to 1.9 kHz. Increasing the
measurement bandwidth is important for many applications of atomic
magnetometery, such as detection of biological magnetic fields from
the heart and the brain \cite{Weis,Xia} and nuclear magnetic
resonance \cite{Savukov,Xu,Ledbetter}, where the bandwidth of
magnetic signals often exceeds the natural bandwidth of the atomic
magnetometers. We operate the magnetometer in a scalar measurement
mode and obtain magnetic field sensitivity of 22 fT/Hz$^{1/2}$, in
agreement with theoretical prediction for the size of spin
projection noise and only a factor of 2 away from the best
sensitivity previously obtained in a scalar magnetometer
\cite{Smullin}. Our measurements are performed with spin
polarization of only 1\%, where back-action evasion is not
necessary. However, our measurement configuration is naturally
back-action evading and thus much higher sensitivity can be expected
in a highly-polarized atomic vapor.

\begin{figure}[tbp]
\includegraphics[width=7.5 cm]{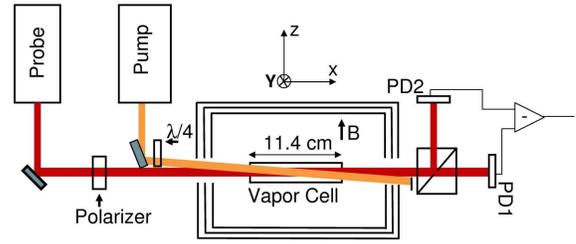}
\caption{Apparatus for high-bandwidth QND scalar magnetometry.}
\label{setup}
\end{figure}

The experimental geometry is shown in Fig.~\ref{setup}. A Pyrex
glass cell 1 cm in diameter and 11.4 cm long contains enriched
$^{87}$Rb and 60 torr of Nitrogen buffer gas and is heated to about
110$^{\circ }$C. The cell
is placed inside multi-layer magnetic shields and a magnetic field of 4.4 $%
\mu$T generated by an ultra-stable current source is directed
perpendicular to the long axis of the cell, giving a Larmor
frequency of 31 kHz. A linearly polarized probe beam detuned from
the D1 line is directed along the length of the cell. Paramagnetic
Faraday rotation induced by the atoms is measured with a balanced
polarimeter.

First we measure the spin-projection noise in an unpolarized vapor.
The power spectrum of the polarimeter output is plotted in Fig.~2,
showing a large noise peak centered at the Larmor frequency.
Detection of alkali-metal spin fluctuations by paramagnetic Faraday
rotation was first demonstrated by Alexandrov \cite{Alexandrov} and
later studied in \cite{Crooker,Crooker1,Kominis}. We obtain a ratio
of the peak spin noise power to the flat photon shot noise
background equal to 22, substantially higher than in previous
experiments. For our conditions  the probe laser is detuned far away
relative to the excited state hyperfine structure and the Lorentzian
and Doppler linewidths, but comparable to the ground-state hyperfine
splitting. The optical rotation angle in this case is proportional
to the vector spin polarization, with negligible contribution from
the tensor polarization \cite{Mathur,Mabuchi}. For the D1 line the
rotation angle is given by
\begin{equation}
\phi =\frac{cr_{e}f_{osc}nl}{(2I+1)}[D(\nu -\nu _{a})\langle
F_{x}^{a}\rangle -D(\nu -\nu _{b})\langle F_{x}^{b}\rangle ],
\end{equation}
where $r_{e}=2.82\times 10^{-13}$ cm is the classical electron radius, $%
f_{osc}=0.34$ is the oscillator strength of the D1 transition in Rb,
$n$ is the vapor density of alkali-metal atoms, $l$ is the length of
the cell along the probe direction, $\langle F_{x}^{a}\rangle $ and
$\langle F_{x}^{b}\rangle $
are the expectation values of the atomic spin $\mathbf{F=I+S}$ in the $%
a=I+1/2$ and $b=I-1/2$ hyperfine states with optical resonance
frequencies of $\nu _{a}$ and $\nu _{b}$. Here $D(\nu )$ is the
dispersion profile given by $D(\nu )=(\nu -\nu _{0})/[(\nu -\nu
_{0})^{2}+(\Delta \nu /2)^{2}]$, where $\Delta \nu $ is the
Lorentzian FWHM due to pressure broadening. For unpolarized,
uncorrelated atoms the r.m.s. spin fluctations are given by
\begin{equation}
\sqrt{\overline{\langle F_{x}^{a}\rangle ^{2}}}=\sqrt{\frac{%
F^{a}(F^{a}+1)(2F^{a}+1)}{6(2I+1)N}},
\end{equation}
where $N$ is the number of atoms being probed, and similar for
$F_{x}^{b}$. Most spin relaxation mechanisms in alkali-metal vapor
affect only the electron spin and hence can introduce correlations
between $F_{x}^{a}$ and $F_{x}^{b}$. However, in the regime when the
Larmor frequency $\omega $ is much larger than the spin relaxation
rate, the transverse spin components of $F^{a}$ and $F^{b}$ precess
in opposite directions, quickly destroying any correlations. Hence
polarization rotation noise from the two terms in Eq.~(1) is not
correlated. The number of atoms participating in the measurement
depends on the transverse intensity profile of the probe beam
$I(y,z)$ and is determined by summing noise variance from different
parts of the beam,
\begin{equation}
N=nl\frac{\left[ \int I(y,z)dydz\right] ^{2}}{\int I(y,z)^{2}dydz}.
\end{equation}
The data in Fig.~2 are very well described by a single Lorentzian
and a constant photon shot noise background. The Lorentzian
half-width of the noise peak is equal to 340 Hz, with about half of
the width due to spin-exchange relaxation and the rest due to
absorption of the probe beam and diffusion. The flat noise
background is in agreement with expected level of photon shot noise.
As discussed in \cite{Crooker,Crooker1,Kominis}, the noise spectrum
in general is a sum of Lorentzians, but the widths of the resonances
for $\langle F_{x}^{a}\rangle $ and $\langle F_{x}^{b}\rangle $ are
similar and the strength of the $\langle F_{x}^{b}\rangle $ signal
is smaller, so it is difficult to distinguish them. There are only
small hints of a deviation from a Lorentzian in the wings of the
noise peak. The total r.m.s. noise $\phi _{rms}$ given by the area
under the peak of the power spectrum is not affected by the shape of
the spectrum. Diffusion of atoms in and out of the probe beam also
distorts the spectrum of the noise from a Lorentzian shape, causing
a sharper peak due to Ramsey narrowing \cite{Xiao}, but it also does
not change $\phi _{rms}$.
\begin{figure}[tbp]
\includegraphics[width=8 cm]{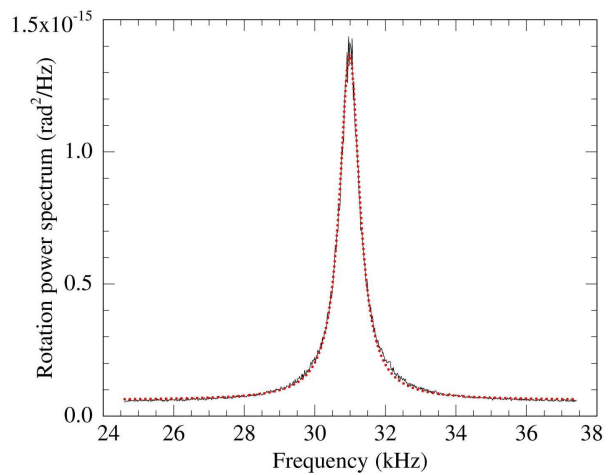}
\caption{ Measured spin noise power spectrum of unpolarized atoms
(solid line) with a fit to a Lorentzian plus flat photon shot noise
background (dashed line). The averaging time for the noise spectrum
is 350 s.} \label{bump}
\end{figure}

For the data presented in Fig.~2 the density of $^{87}$Rb atoms is
equal $n=8.7\times 10^{12}$ cm$^{-3}$,  determined by a measurement
of the transverse spin relaxation time due to spin exchange
collisions. The intensity profile of the probe beam was measured in
both directions by the scanning edge technique and had effective
dimensions of $3.8\times 4.5$ mm$^{2}$. We find that the effective
number of atoms being probed is $N=$ $1.7\times 10^{13}$.
The probe laser is detuned by 19 GHz from the $F=2$ state and 25.8 GHz from $%
F=1$ state, much larger than the Lorentzian optical FWHM $\Delta \nu
=$1.42 GHz. According to Eq (1), this gives $\phi
_{rms}^{th}=1.07\times 10^{-6}$ rad. Experimentally, the area under
the spin noise peak is equal $\phi _{rms}^{\exp }=$ $1.19\times
10^{-6}$ rad. The agreement at the 10\% level is quite good given
the uncertainty in the number of atoms participating in the
measurement.  Using a different cell with natural abundance of Rb
isotopes we verified experimentally that the ratio of spin noise for
$^{87}$Rb and $^{85}$Rb isotopes, which have different nuclear
spins, is consistent with our analysis within a few percent. In a
previous detailed study of atomic spin noise \cite{Crooker1} the
overall level of spin noise was off by a factor of 2 from
predictions.

A scalar atomic magnetometer measuring the absolute value the
magnetic field is realized using a circularly polarized pump beam
propagating nearly parallel to the probe beam but missing the
photodetectors. We use Bell-Bloom excitation \cite{BellBloom} of the
spin precession around $B_{z}$ field by sinusoidally modulating the
current in a DFB laser used to generate the pumping light at the
Larmor frequency.  The polarimeter signal is directly digitized
using a fast, high resolution A/D card and lock-in demodulation of
the data is implemented at the analysis stage. The noise spectrum of
the out-of-phase component of the lock-in signal, proportional to
small variations of $B_z$ field, is shown in Fig.~3 with and without
the pump beam. The intensity of the pump beam is increased until the
noise level just starts to increase due to technical noise sources
and is equal to about 100 $\mu$W. The spin polarization of the
atomic vapor, determined from the amplitude of the oscillating
rotation signal, is equal to 1.0\%. Discrete noise peaks from
magnetic interference can be seen for polarized atoms. The magnetic
field sensitivity of the rotation signal is calibrated by applying a
known modulation to $B_{z}$ field at various frequencies. We show in
Fig.~3 with filled circles the expected magnetometer signal for a 22
fT$_{rms}$ oscillating magnetic field as a function of frequency. It
can be seen that at higher frequencies the response of the
magnetometer drops, but the noise level decreases as well, so the
magnetometer retains its sensitivity up to much higher frequencies
than the resonance linewidth.
\begin{figure}[tbp]
\includegraphics[width=7.5 cm]{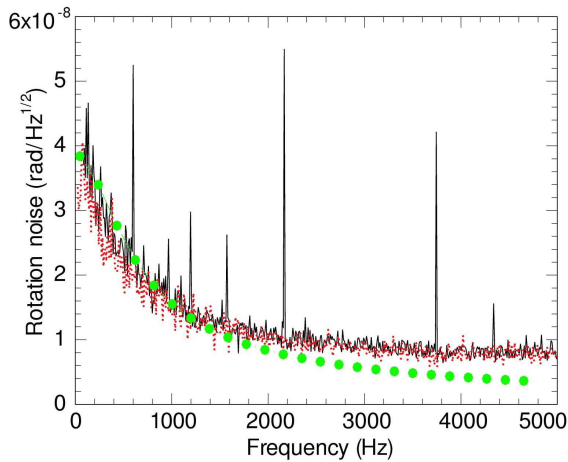}
\caption{Lock-in output amplitude noise spectrum for polarized atoms
(solid line) and unpolarized atoms (dashed line).  The signal
expected in the magnetometer for a 22 fT$_{\rm rms}$ magnetic field
is shown with dots. The spikes in the noise spectrum for polarized
atoms are from technical magnetic noise. } \label{lockin}
\end{figure}

In fact, the sensitivity remains constant as long as the noise
spectrum is dominated by spin noise. This can be easily seen in the
simpler case of a spin-1/2 system when the magnetic resonance is
described by Bloch equations. One can show that the absolute value
of the magnetometer signal in response to an oscillating $B_{z}$
field decreases with frequency as the square root of a Lorentzian
with a width given by the inverse of the spin coherence time
$T_{2}$,
\begin{equation}
S(f)=S_{0}/[1+(2\pi fT_{2})^{2}]^{1/2}.  \label{Resp}
\end{equation}
The shape of the spin noise spectrum is also described by the square
root of a Lorentzian with the same width \cite{Braun}. Hence the
signal and the noise decrease simultaneously, maintaining constant
sensitivity. In Fig.~4 we show the magnetic field sensitivity of the
magnetometer as a function of frequency by dividing the noise
spectrum by the response curve. The sensitivity remains nearly
constant up to 2 kHz, while the resonance linewidth of the
magnetometer is equal to 420 Hz in this case. The increase in the
bandwidth is a direct result of the non-white nature of the spin
noise that is realized with QND measurements which preserve temporal
correlations of the spin expectation value. In contrast, if the spin
polarization is monitored using optical absorption instead of
Faraday rotation, or if the optical density of the vapor on
resonance is less than one, the noise spectrum would be white. We
show in Fig.~4 the sensitivity that would be obtained with such a
flat noise spectrum for comparison. This demonstrates the role of
quantum-non-demolition measurements, they do not improve the
performance at low frequencies but increase the bandwidth of
measurements without any penalty in sensitivity. The necessary
condition for  bandwidth increase can be written as $(\Delta
J_x)^2<J/2$  in order to resolve temporal spin correlations. This is
a weaker condition than $(\Delta J_x)^2<\langle J_z \rangle/2$ that
usually defines spin squeezing \cite{Wineland}.

\begin{figure}[tbp]
\includegraphics[width=8 cm]{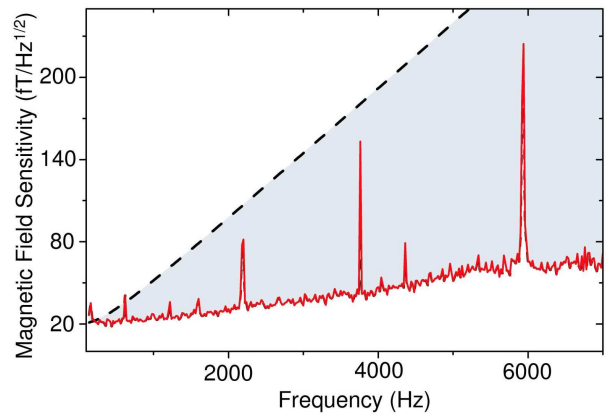}
\caption{Experimentally observed magnetic field noise spectral
density corrected for the frequency response of the magnetometer
(solid line).  The spikes are due to narrow-band magnetic noise.
Dashed line shows expected magnetometer sensitivity for a demolition
measurement assuming a flat noise spectrum. The shaded area
represents improvement in the magnetometer sensitivity at high
frequencies as a result of QND measurements. For QND measurements
the sensitivity decreases by $\sqrt{2}$ at 1.9 kHz. }
\label{lockinBW}
\end{figure}

In general, one can define the measurement bandwidth of the magnetometer as
the frequency at which the sensitivity of the magnetometer drops by a factor
of $1/\sqrt{2}$ from its value at low frequency. For a square root of
Lorentzian frequency response and a flat noise spectrum $f_{BW}=1/(2\pi
T_{2}).$ Using electronic feedback or self-oscillating operation it is
possible to obtain a flat frequency response, however, even in this case the
sensitivity of the magnetometer becomes worse at frequencies higher than $%
1/T_{2}$ because of noise induced by the feedback \cite{Schwindt}.
For a quantum non-demolition paramagnetic Faraday rotation
measurement with a far-detuned probe laser, the spectral noise
density can be written as
\begin{equation}
\phi _{n}\left( f\right) =\frac{1}{\sqrt{2\Phi  }}\left\{ \frac{1}{\eta}+\frac{%
N_{ab}\Gamma _{pr}T_{2}\beta }{1+[2\pi (f-f_{0})T_{2}]^{2}}\right\}
^{1/2}, \label{Noise}
\end{equation}
where $\Phi $ is the probe laser photon flux, $\eta $ is the quantum
efficiency of the photodetectos,  $N_{ab}$ is the number of
absorption lengths (optical density) of the atomic sample on
resonance, $\Gamma _{pr}$ is the optical pumping rate of the probe
laser and $\beta $ is a factor of order unity depending on the
nuclear spin and polarization of the atomic ensemble ($\beta =1$ for
$I=0$). Using Eq. (\ref{Resp}) and (\ref {Noise}) it's easy to show
that the bandwidth of the magnetometer with quantum-non-demolition
measurements is given by $f_{BW}=(\eta N_{ab}\Gamma _{pr}T_{2}\beta
+1)^{1/2}/(2\pi T_{2})$. In the regime where the spin relaxation
rate is dominated by the optical pumping rate of the probe laser,
$\Gamma _{pr}T_{2}\simeq 1$, the bandwidth is increased by a factor
on the order of  the square root of the optical density on resonance
$N_{ab}^{1/2}$.

For our conditions the  back-action of the probe beam is not
significant because of low spin polarization. However, our
experimental arrangement is naturally back-action evading and can be
used to generate conditional spin squeezing. In the regime of far
deturning the back-action of the probe beam is due to the light
shift created by the quantum fluctuations of the circular
polarization of the probe beam and is equivalent to a fictitious
magnetic field parallel to the propagation direction of the probe
beam \cite{Mathur}. Our probe beam is directed perpendicular to a
large static magnetic field and thus the light shift fluctuations
only contribute in second order to the absolute value of the
magnetic field measured by the scalar magnetometer.  A fluctuating
magnetic field along the $x$ direction will generate a small $z$
component of the polarization that is not directly measured.
Therefore this arrangement can be used to generate conditional spin
squeezing in a highly polarized vapor \cite{Kuzmich}.

The magnetic field sensitivity of 22 fT/Hz$^{1/2}$ obtained in this
experiment is the best measured sensitivity with a single-channel
scalar magnetometer and within a factor of 2 of the best measured
sensitivity obtained in a gradiometer arrangement \cite{Smullin}.
Scalar atomic magnetometers have lagged in sensitivity compared to
other types of atomic magnetometers \cite{MagNatRev} because of
spin-exchange broadening. The measurement volume used in our sensor
is about 2 cm$^{3}$ and the spin polarization is equal to 1\%. The
sensitivity of the magnetometer will be optimized if the intensity
of the pumping light is increased to obtain spin polarization of
50\% with a factor of 2 broadening of the magnetic resonance.  The
signal will increase by a factor of 50, while the spin-projection
noise will not change appreciably, resulting in a magnetic field
sensitivity of about 0.6~fT/Hz$^{1/2}$, consistent with
spin-exchange limited sensitivity for alkali-metal magnetometers
\cite{Smullin}. Alternatively, operating in the regime of low spin
polarization can be advantageous if it is desired to minimize the
heading errors of the magnetometer which depend on the degree of
polarization \cite{Seltzer}.

In conclusion, we investigated operation of an atomic magnetometer
in the spin-projection noise limited regime.  The magnitude of the
spin noise is in good agreement with theory. We demonstrated an
increase in the magnetometer bandwidth by a factor of 4 using
quantum non-demolition measurements. Such increase of the
measurement bandwidth without loss of sensitivity is important for
many practical applications of atomic magnetometery, such as
detection of NMR and biological fields. Similar QND measurements can
also be used to increase the bandwidth of magnetometers based on
nuclear spins with very long spin coherence times and the update
rate of atomic clocks based on very narrow transitions. This work
was supported by an ONR MURI award.

\begin{thebibliography}{24}
\expandafter\ifx\csname natexlab\endcsname\relax\def\natexlab#1{#1}\fi
\expandafter\ifx\csname bibnamefont\endcsname\relax
  \def\bibnamefont#1{#1}\fi
\expandafter\ifx\csname bibfnamefont\endcsname\relax
  \def\bibfnamefont#1{#1}\fi
\expandafter\ifx\csname citenamefont\endcsname\relax
  \def\citenamefont#1{#1}\fi
\expandafter\ifx\csname url\endcsname\relax
  \def\url#1{\texttt{#1}}\fi
\expandafter\ifx\csname urlprefix\endcsname\relax\def\urlprefix{URL }\fi
\providecommand{\bibinfo}[2]{#2}
\providecommand{\eprint}[2][]{\url{#2}}

\bibitem[{\citenamefont{Wineland et~al.}(1992)\citenamefont{Wineland,
  Bollinger, Itano, Moore, and Heinzen}}]{Wineland}
\bibinfo{author}{\bibfnamefont{D.~J.} \bibnamefont{Wineland}},
  \bibinfo{author}{\bibfnamefont{J.~J.} \bibnamefont{Bollinger}},
  \bibinfo{author}{\bibfnamefont{W.~M.} \bibnamefont{Itano}},
  \bibinfo{author}{\bibfnamefont{F.~L.} \bibnamefont{Moore}}, \bibnamefont{and}
  \bibinfo{author}{\bibfnamefont{D.~J.} \bibnamefont{Heinzen}},
  \bibinfo{journal}{Phys. Rev. A} \textbf{\bibinfo{volume}{46}},
  \bibinfo{pages}{R6797} (\bibinfo{year}{1992}).

\bibitem[{\citenamefont{Huelga et~al.}(1997)\citenamefont{Huelga, Macchiavello,
  Pellizzari, Ekert, Plenio, and Cirac}}]{Huelga}
\bibinfo{author}{\bibfnamefont{S.~F.} \bibnamefont{Huelga}},
  \bibinfo{author}{\bibfnamefont{C.}~\bibnamefont{Macchiavello}},
  \bibinfo{author}{\bibfnamefont{T.}~\bibnamefont{Pellizzari}},
  \bibinfo{author}{\bibfnamefont{A.~K.} \bibnamefont{Ekert}},
  \bibinfo{author}{\bibfnamefont{M.~B.} \bibnamefont{Plenio}},
  \bibnamefont{and} \bibinfo{author}{\bibfnamefont{J.~I.} \bibnamefont{Cirac}},
  \bibinfo{journal}{Phys. Rev. Lett.} \textbf{\bibinfo{volume}{79}},
  \bibinfo{pages}{3865} (\bibinfo{year}{1997}).

\bibitem[{\citenamefont{Andr\'e et~al.}(2004)\citenamefont{Andr\'e,
  S\o{}rensen, and Lukin}}]{Andre}
\bibinfo{author}{\bibfnamefont{A.}~\bibnamefont{Andr\'e}},
  \bibinfo{author}{\bibfnamefont{A.~S.} \bibnamefont{S\o{}rensen}},
  \bibnamefont{and} \bibinfo{author}{\bibfnamefont{M.~D.} \bibnamefont{Lukin}},
  \bibinfo{journal}{Phys. Rev. Lett.} \textbf{\bibinfo{volume}{92}},
  \bibinfo{pages}{230801} (\bibinfo{year}{2004}).

\bibitem[{\citenamefont{Auzinsh et~al.}(2004)\citenamefont{Auzinsh, Budker,
  Kimball, Rochester, Stalnaker, Sushkov, and Yashchuk}}]{Auzinsh}
\bibinfo{author}{\bibfnamefont{M.}~\bibnamefont{Auzinsh}},
  \bibinfo{author}{\bibfnamefont{D.}~\bibnamefont{Budker}},
  \bibinfo{author}{\bibfnamefont{D.~F.} \bibnamefont{Kimball}},
  \bibinfo{author}{\bibfnamefont{S.~M.} \bibnamefont{Rochester}},
  \bibinfo{author}{\bibfnamefont{J.~E.} \bibnamefont{Stalnaker}},
  \bibinfo{author}{\bibfnamefont{A.~O.} \bibnamefont{Sushkov}},
  \bibnamefont{and} \bibinfo{author}{\bibfnamefont{V.~V.}
  \bibnamefont{Yashchuk}}, \bibinfo{journal}{Phys. Rev. Lett.}
  \textbf{\bibinfo{volume}{93}}, \bibinfo{pages}{173002}
  (\bibinfo{year}{2004}).

\bibitem[{\citenamefont{Budker and Romalis}(2007)}]{MagNatRev}
\bibinfo{author}{\bibfnamefont{D.}~\bibnamefont{Budker}} \bibnamefont{and}
  \bibinfo{author}{\bibfnamefont{M.}~\bibnamefont{Romalis}},
  \bibinfo{journal}{Nature Physics} \textbf{\bibinfo{volume}{3}},
  \bibinfo{pages}{227} (\bibinfo{year}{2007}).

\bibitem[{\citenamefont{Takahashi et~al.}(1999)\citenamefont{Takahashi, Honda,
  Tanaka, Toyoda, Ishikawa, and Yabuzaki}}]{Takahashi}
\bibinfo{author}{\bibfnamefont{Y.}~\bibnamefont{Takahashi}},
  \bibinfo{author}{\bibfnamefont{K.}~\bibnamefont{Honda}},
  \bibinfo{author}{\bibfnamefont{N.}~\bibnamefont{Tanaka}},
  \bibinfo{author}{\bibfnamefont{K.}~\bibnamefont{Toyoda}},
  \bibinfo{author}{\bibfnamefont{K.}~\bibnamefont{Ishikawa}}, \bibnamefont{and}
  \bibinfo{author}{\bibfnamefont{T.}~\bibnamefont{Yabuzaki}},
  \bibinfo{journal}{Phys. Rev. A} \textbf{\bibinfo{volume}{60}},
  \bibinfo{pages}{4974} (\bibinfo{year}{1999}).

\bibitem[{\citenamefont{Bison et~al.}(2003)\citenamefont{Bison, Wynands, and
  Weis}}]{Weis}
\bibinfo{author}{\bibfnamefont{G.}~\bibnamefont{Bison}},
  \bibinfo{author}{\bibfnamefont{R.}~\bibnamefont{Wynands}}, \bibnamefont{and}
  \bibinfo{author}{\bibfnamefont{A.}~\bibnamefont{Weis}},
  \bibinfo{journal}{Optics Express} \textbf{\bibinfo{volume}{11}},
  \bibinfo{pages}{904} (\bibinfo{year}{2003}).

\bibitem[{\citenamefont{Xia et~al.}(2006)\citenamefont{Xia, Baranga, Hoffman,
  and Romalis}}]{Xia}
\bibinfo{author}{\bibfnamefont{H.}~\bibnamefont{Xia}},
  \bibinfo{author}{\bibfnamefont{A.~B.-A.} \bibnamefont{Baranga}},
  \bibinfo{author}{\bibfnamefont{D.}~\bibnamefont{Hoffman}}, \bibnamefont{and}
  \bibinfo{author}{\bibfnamefont{M.~V.} \bibnamefont{Romalis}},
  \bibinfo{journal}{Appl. Phys. Lett.} \textbf{\bibinfo{volume}{89}}
  (\bibinfo{year}{2006}).

\bibitem[{\citenamefont{Savukov and Romalis}(2005)}]{Savukov}
\bibinfo{author}{\bibfnamefont{I.~M.} \bibnamefont{Savukov}} \bibnamefont{and}
  \bibinfo{author}{\bibfnamefont{M.~V.} \bibnamefont{Romalis}},
  \bibinfo{journal}{Phys. Rev. Lett.} \textbf{\bibinfo{volume}{94}},
  \bibinfo{pages}{123001} (\bibinfo{year}{2005}).

\bibitem[{\citenamefont{Xu et~al.}(2006)\citenamefont{Xu, Yashchuk, Donaldson,
  Rochester, Budker, and Pines}}]{Xu}
\bibinfo{author}{\bibfnamefont{S.}~\bibnamefont{Xu}},
  \bibinfo{author}{\bibfnamefont{V.~V.} \bibnamefont{Yashchuk}},
  \bibinfo{author}{\bibfnamefont{M.~H.} \bibnamefont{Donaldson}},
  \bibinfo{author}{\bibfnamefont{S.~M.} \bibnamefont{Rochester}},
  \bibinfo{author}{\bibfnamefont{D.}~\bibnamefont{Budker}}, \bibnamefont{and}
  \bibinfo{author}{\bibfnamefont{A.}~\bibnamefont{Pines}},
  \bibinfo{journal}{Proc. Nat. Acad. Sci.} \textbf{\bibinfo{volume}{103}},
  \bibinfo{pages}{12668} (\bibinfo{year}{2006}).

\bibitem[{\citenamefont{Ledbetter et~al.}(2008)\citenamefont{Ledbetter,
  Savukov, Budker, Shah, Knappe, Kitching, Michalak, Xu, and
  Pines}}]{Ledbetter}
\bibinfo{author}{\bibfnamefont{M.~P.} \bibnamefont{Ledbetter}},
  \bibinfo{author}{\bibfnamefont{I.~M.} \bibnamefont{Savukov}},
  \bibinfo{author}{\bibfnamefont{D.}~\bibnamefont{Budker}},
  \bibinfo{author}{\bibfnamefont{V.}~\bibnamefont{Shah}},
  \bibinfo{author}{\bibfnamefont{S.}~\bibnamefont{Knappe}},
  \bibinfo{author}{\bibfnamefont{J.}~\bibnamefont{Kitching}},
  \bibinfo{author}{\bibfnamefont{D.~J.} \bibnamefont{Michalak}},
  \bibinfo{author}{\bibfnamefont{S.}~\bibnamefont{Xu}}, \bibnamefont{and}
  \bibinfo{author}{\bibfnamefont{A.}~\bibnamefont{Pines}},
  \bibinfo{journal}{Proc. Nat. Acad. Sci.} \textbf{\bibinfo{volume}{105}},
  \bibinfo{pages}{2286} (\bibinfo{year}{2008}).

\bibitem[{\citenamefont{Smullin et~al.}(2009)\citenamefont{Smullin, Savukov,
  Vasilakis, Ghosh, and Romalis}}]{Smullin}
\bibinfo{author}{\bibfnamefont{S.~J.} \bibnamefont{Smullin}},
  \bibinfo{author}{\bibfnamefont{I.~M.} \bibnamefont{Savukov}},
  \bibinfo{author}{\bibfnamefont{G.}~\bibnamefont{Vasilakis}},
  \bibinfo{author}{\bibfnamefont{R.~K.} \bibnamefont{Ghosh}}, \bibnamefont{and}
  \bibinfo{author}{\bibfnamefont{M.~V.} \bibnamefont{Romalis}},
  \bibinfo{journal}{Phys. Rev. A} \textbf{\bibinfo{volume}{80}},
  \bibinfo{pages}{033420} (\bibinfo{year}{2009}).

\bibitem[{\citenamefont{Alexsandrov and Zapassky}(1981)}]{Alexandrov}
\bibinfo{author}{\bibfnamefont{E.}~\bibnamefont{Alexsandrov}} \bibnamefont{and}
  \bibinfo{author}{\bibfnamefont{V.}~\bibnamefont{Zapassky}},
  \bibinfo{journal}{Zh. Eksp. Teor. Fiz.} \textbf{\bibinfo{volume}{81}},
  \bibinfo{pages}{132} (\bibinfo{year}{1981}).

\bibitem[{\citenamefont{Crooker et~al.}(2004)\citenamefont{Crooker, Rickel,
  Balatsky, and Smith}}]{Crooker}
\bibinfo{author}{\bibfnamefont{S.}~\bibnamefont{Crooker}},
  \bibinfo{author}{\bibfnamefont{D.}~\bibnamefont{Rickel}},
  \bibinfo{author}{\bibfnamefont{A.}~\bibnamefont{Balatsky}}, \bibnamefont{and}
  \bibinfo{author}{\bibfnamefont{D.}~\bibnamefont{Smith}},
  \bibinfo{journal}{Nature} \textbf{\bibinfo{volume}{431}}, \bibinfo{pages}{49}
  (\bibinfo{year}{2004}).

\bibitem[{\citenamefont{Mihaila et~al.}(2006)\citenamefont{Mihaila, Crooker,
  Rickel, Blagoev, Littlewood, and Smith}}]{Crooker1}
\bibinfo{author}{\bibfnamefont{B.}~\bibnamefont{Mihaila}},
  \bibinfo{author}{\bibfnamefont{S.~A.} \bibnamefont{Crooker}},
  \bibinfo{author}{\bibfnamefont{D.~G.} \bibnamefont{Rickel}},
  \bibinfo{author}{\bibfnamefont{K.~B.} \bibnamefont{Blagoev}},
  \bibinfo{author}{\bibfnamefont{P.~B.} \bibnamefont{Littlewood}},
  \bibnamefont{and} \bibinfo{author}{\bibfnamefont{D.~L.} \bibnamefont{Smith}},
  \bibinfo{journal}{Phys. Rev. A} \textbf{\bibinfo{volume}{74}},
  \bibinfo{pages}{043819} (\bibinfo{year}{2006}).

\bibitem[{\citenamefont{Katsoprinakis et~al.}(2007)\citenamefont{Katsoprinakis,
  Dellis, and Kominis}}]{Kominis}
\bibinfo{author}{\bibfnamefont{G.~E.} \bibnamefont{Katsoprinakis}},
  \bibinfo{author}{\bibfnamefont{A.~T.} \bibnamefont{Dellis}},
  \bibnamefont{and} \bibinfo{author}{\bibfnamefont{I.~K.}
  \bibnamefont{Kominis}}, \bibinfo{journal}{Phys. Rev. A}
  \textbf{\bibinfo{volume}{75}}, \bibinfo{pages}{042502}
  (\bibinfo{year}{2007}).

\bibitem[{\citenamefont{Mathur et~al.}(1968)\citenamefont{Mathur, Tang, and
  Happer}}]{Mathur}
\bibinfo{author}{\bibfnamefont{B.~S.} \bibnamefont{Mathur}},
  \bibinfo{author}{\bibfnamefont{H.}~\bibnamefont{Tang}}, \bibnamefont{and}
  \bibinfo{author}{\bibfnamefont{W.}~\bibnamefont{Happer}},
  \bibinfo{journal}{Phys. Rev.} \textbf{\bibinfo{volume}{171}},
  \bibinfo{pages}{11} (\bibinfo{year}{1968}).

\bibitem[{\citenamefont{Geremia et~al.}(2006)\citenamefont{Geremia, Stockton,
  and Mabuchi}}]{Mabuchi}
\bibinfo{author}{\bibfnamefont{J.~M.} \bibnamefont{Geremia}},
  \bibinfo{author}{\bibfnamefont{J.~K.} \bibnamefont{Stockton}},
  \bibnamefont{and} \bibinfo{author}{\bibfnamefont{H.}~\bibnamefont{Mabuchi}},
  \bibinfo{journal}{Phys. Rev. A} \textbf{\bibinfo{volume}{73}},
  \bibinfo{pages}{042112} (\bibinfo{year}{2006}).

\bibitem[{\citenamefont{Xiao et~al.}(2006)\citenamefont{Xiao, Novikova,
  Phillips, and Walsworth}}]{Xiao}
\bibinfo{author}{\bibfnamefont{Y.}~\bibnamefont{Xiao}},
  \bibinfo{author}{\bibfnamefont{I.}~\bibnamefont{Novikova}},
  \bibinfo{author}{\bibfnamefont{D.~F.} \bibnamefont{Phillips}},
  \bibnamefont{and} \bibinfo{author}{\bibfnamefont{R.~L.}
  \bibnamefont{Walsworth}}, \bibinfo{journal}{Phys. Rev. Lett.}
  \textbf{\bibinfo{volume}{96}}, \bibinfo{pages}{043601}
  (\bibinfo{year}{2006}).

\bibitem[{\citenamefont{Bell and Bloom}(1961)}]{BellBloom}
\bibinfo{author}{\bibfnamefont{W.~E.} \bibnamefont{Bell}} \bibnamefont{and}
  \bibinfo{author}{\bibfnamefont{A.~L.} \bibnamefont{Bloom}},
  \bibinfo{journal}{Phys. Rev. Lett.} \textbf{\bibinfo{volume}{6}},
  \bibinfo{pages}{280} (\bibinfo{year}{1961}).

\bibitem[{\citenamefont{Braun and K\"onig}(2007)}]{Braun}
\bibinfo{author}{\bibfnamefont{M.}~\bibnamefont{Braun}} \bibnamefont{and}
  \bibinfo{author}{\bibfnamefont{J.}~\bibnamefont{K\"onig}},
  \bibinfo{journal}{Phys. Rev. B} \textbf{\bibinfo{volume}{75}},
  \bibinfo{pages}{085310} (\bibinfo{year}{2007}).

\bibitem[{\citenamefont{Schwindt et~al.}(2005)\citenamefont{Schwindt, Hollberg,
  and Kitching}}]{Schwindt}
\bibinfo{author}{\bibfnamefont{P.}~\bibnamefont{Schwindt}},
  \bibinfo{author}{\bibfnamefont{L.}~\bibnamefont{Hollberg}}, \bibnamefont{and}
  \bibinfo{author}{\bibfnamefont{J.}~\bibnamefont{Kitching}},
  \bibinfo{journal}{Rev.Sci. Instrum.} \textbf{\bibinfo{volume}{76}}
  (\bibinfo{year}{2005}).

\bibitem[{\citenamefont{Kuzmich et~al.}(2000)\citenamefont{Kuzmich, Mandel, and
  Bigelow}}]{Kuzmich}
\bibinfo{author}{\bibfnamefont{A.}~\bibnamefont{Kuzmich}},
  \bibinfo{author}{\bibfnamefont{L.}~\bibnamefont{Mandel}}, \bibnamefont{and}
  \bibinfo{author}{\bibfnamefont{N.~P.} \bibnamefont{Bigelow}},
  \bibinfo{journal}{Phys. Rev. Lett.} \textbf{\bibinfo{volume}{85}},
  \bibinfo{pages}{1594} (\bibinfo{year}{2000}).

\bibitem[{\citenamefont{Seltzer et~al.}(2007)\citenamefont{Seltzer, Meares, and
  Romalis}}]{Seltzer}
\bibinfo{author}{\bibfnamefont{S.~J.} \bibnamefont{Seltzer}},
  \bibinfo{author}{\bibfnamefont{P.~J.} \bibnamefont{Meares}},
  \bibnamefont{and} \bibinfo{author}{\bibfnamefont{M.~V.}
  \bibnamefont{Romalis}}, \bibinfo{journal}{Phys. Rev. A}
  \textbf{\bibinfo{volume}{75}}, \bibinfo{pages}{051407}
  (\bibinfo{year}{2007}).

\end{thebibliography}

\end{document}